\documentclass[iop,revtex4]{emulateapj}
\newcommand{\mytilde}{\raise.19ex\hbox{$\scriptstyle\sim$}}
\usepackage{hyperref}
\usepackage{xcolor}
\usepackage{makecell}
\begin{document}
\newcommand{\kms}{\ensuremath{\mathrm{km\,s}^{-1}}}

\title{Baryonic feedback measurement from KV450 cosmic shear analysis}
\author{MIJIN YOON\altaffilmark{1,2} and M. JAMES JEE\altaffilmark{2,3}}

\altaffiltext{1}{Ruhr-University Bochum, Astronomical Institute, German Centre for Cosmological Lensing, Universitätsstr. 150, 44801 Bochum, Germany; yoon@astro.rub.de}
\altaffiltext{2}{Department of Astronomy, Yonsei University, Yonsei-ro 50, Seoul, Korea; mjyoon@yonsei.ac.kr, mkjee@yonsei.ac.kr}
\altaffiltext{3}{Department of Physics, University of California, Davis, California, USA}
\keywords{baryonic feedback --- cosmology --- gravitational lensing: weak --- astrophysics: observations --- large-scale structure of the Universe}

\begin{abstract}
While baryonic feedback is one of the most important astrophysical systematics that we need to address in order to achieve precision cosmology, few weak lensing studies have directly measured its impact on the matter power spectrum. We report measurement of the baryonic feedback parameter with the constraints on its lower and upper limits from cosmic shear.
We use the public data from the Kilo-Degree Survey and the VISTA Kilo-Degree Infrared Galaxy Survey spanning 450 deg$^2$.
Estimating both cosmological and feedback parameters simultaneously, we
obtain
$A_{\rm b}=1.01_{-0.85}^{+0.80}$,
which shows a consistency with the dark matter-only (DMO) case at the $\mytilde1.2~\sigma$ level and a tendency toward positive feedback; the $A_{\rm b}=0$ ($0.81$) value corresponds to the DMO (OWLS AGN) case.
Despite this full constraint of the feedback parameter,
our $S_8~(\equiv \sigma_8 \sqrt{\Omega_m / 0.3})$ measurement ($0.739^{+0.036}_{- 0.035}$) shifts by
only $\mytilde6$\% of the statistical error, 
compared to the previous measurement.
When we assume the flat $\Lambda$CDM cosmology favored by the Nine-Year Wilkinson Microwave Anisotropy Probe (Planck) result, the feedback parameter is constrained to be $A_{\rm b}=1.21_{-0.54}^{+0.61}$
($1.60_{-0.52}^{+0.53}$), which excludes the DMO  case at the $\mytilde2.2~\sigma$ ($\mytilde3.1~\sigma$)  level. 

\end{abstract}

\section{Introduction}
Weak gravitational lensing (WL) is one of the most powerful probes of dark energy. Stage IV WL projects, such as the Vera C. Rubin Observatory \citep{2019ApJ...873..111I}, Euclid \citep{2011arXiv1110.3193L}, Nancy Grace Roman Space Telescope \citep{2015arXiv150303757S}, etc., are expected to start their missions in the current decade by measuring subtle shape distortions from billions of distant galaxies to pursue this goal.
Certainly, the superb statistical power from these Stage IV experiments will enable the so-called precision cosmology era only if we manage to control their systematic errors below or comparable to the statistical ones.

Accurate determination of the change in the total matter power spectrum (PS) due to baryon physics with respect to the dark matter-only (DMO) case is one of the most difficult challenges in theoretical systematics. Unlike cold dark matter, which is believed to interact only through gravity, it is difficult to perform robust numerical simulations with baryonic feedback because its impact is dominated by so-called sub-grid physics. 
Fortunately, there have been positive developments on this lately, helped by more efforts in observations, calibration strategies, simulation cost reduction, etc., but there are clear degeneracies in sub-grid physics that need to be quantified.
Currently, non-negligible discrepancies in baryonic feedback prediction among different state-of-the-art simulations exist.

The requirement of the matter PS prediction accuracy in the precision cosmology era is very stringent. \cite{2005APh....23..369H} estimated that an accuracy of 1-2\% is needed for the $k$ range $0.1~h~\rm  Mpc^{-1}$ $\lesssim k$ $\lesssim 10~h~\rm  Mpc^{-1}$ in order to fully utilize the statistical power of Stage IV data.
When photometric redshift uncertainty is included,
a much tighter (0.5\%) requirement over a larger $k$ range ($0.01~h\rm~Mpc^{-1}$ $\lesssim k$ $\lesssim  5~h\rm~ Mpc^{-1}$) is suggested by
\cite{2012JCAP...04..034H}.

The traditional approach to addressing the baryonic effect in cosmological WL analysis is to simply truncate the observed correlation function/PS on small scales  \citep[see][for review]{2019OJAp....2E...4C} or to employ  modified estimators \citep[e.g. COSEBIs,][]{Asgari2020} that filter out the part heavily influenced by baryonic physics.
The obvious drawback of these methods are non-negligible reduction of the statistical power particularly because the WL signal-to-noise ratio is high on the nonlinear (thus small) scales.
Also, the absence of the consensus on the optimal baryonic feedback recipe provides ambiguity in determining the exact cutoff scale.
\defcitealias{2015MNRAS.450.1212H}{HD15}

Alternatively, recent studies suggest empirical modeling of the matter PS modification based on hydrodynamical cosmological simulations \citep[e.g.,][]{2011MNRAS.417.2020S,2013MNRAS.434..148S}. Although no consensus is present regarding the exact feedback strength and scale, one can parameterize the effect by recognizing the patterns in the matter PS modification in different simulations.
\cite{2015MNRAS.454.2451E} propose the principal component analysis (PCA) as a method to recognize the pattern. 
One can then either discard the components sensitive to the baryonic feedback or
use them to model the effect.
The potential weakness of this approach is that the resulting principal components are sensitive to the choice of the sample. In other words, unless the sample includes a sufficiently broad range of feedback scenarios, the performance can be non-negligibly compromised \citep{2018ApJ...863..173M}. Another approach is to simply parameterize the deviation
of the matter PS from the DMO PS in a model-independent way. For example, \citet[][hereafter HD15]{2015MNRAS.450.1212H} demonstrate that description of the PS ratio variation for  OverWhelmingly Large cosmological hydrodynamical Simulations
\citep[OWLS;][]{2010MNRAS.402.1536S,2011MNRAS.415.3649V} with 15 parameters provides $<5$\% precision. 
\cite{2018MNRAS.480.3962C} show that
their 4-parameter description is 
adequate to provide a good ($<5$\%) fit to Horizon-AGN by using the baryonic correction model \citep{2015JCAP...12..049S}.
Recently, van Daalen et al. (2020) present an empirical model based on numerical simulations with a wide range of feedback
models, which requires only a single parameter.

\defcitealias{2015MNRAS.454.1958M}{M15}

A halo model-based approach is introduced by \citet[][hereafter M15]{2015MNRAS.454.1958M}, who propose to model the modification of the matter PS with the halo properties affected by baryonic feedback.
The authors note that the two parameters of their halo model, which are linearly related with each other, are sensitive to different feedback scenarios in OWLS. Thus, they propose a method to model the feedback with a single parameter. Although subject to further analysis with a broader range of feedback cases, this provides a convenient formalism to characterize the baryonic feedback effect, in particular, for
the current Stage III WL studies, which do not yet provide sufficient statistical power to discriminate subtle differences in various feedback scenarios.

Also, an hybrid approach modifying DMO simulation results with baryonic halo properties has been suggested \citep{Schneider_2019}. This ``baryonification" method applies small shifts to the particles in the DMO halos to effect the baryonic feedback.  Because baryonic effects are empirically parametrized, the approach enables fast realizations of many nonlinear
cosmic density fields with varying baryonic parameters. 

Current cosmological hydrodynamics simulations calibrate their sub-grid physics recipes by ensuring that the results can reproduce some observational statistics (e.g., Schaye et al. 2015; McCarthy et al. 2017; Pillepich et al. 2017; Springel et al. 2017) such as scaling relations, gas fractions, etc. However, because the recipes are degenerate, the calibration parameters cannot be determined uniquely. The large discrepancy seen in the prediction of the matter power spectrum among the different simulations clearly demonstrates the current limitation.
Recently, a novel approach was proposed by Debackere et al. (2020), who measured the power suppression due to baryonic feedback based on the empirical approach combining X-ray hot gas observations and Halo Occupation Distribution (HOD) modeling. Compared to the previous simulation-based approach, there is a clear merit for this approach because it does not rely on the aforementioned subgrid physics. Nevertheless, this study too has to rely on a number of astrophysical/cosmological assumptions including HOD models, matter density profiles, X-ray emissivity, cluster/group mass function, galaxy bias, etc.

\defcitealias{2019ApJ...870..111Y}{Y19} 

One critical cross-check in the baryonic feedback study is a measurement of the feedback impact from the shape of the matter power spectrum, which is directly probed by cosmic shear. Although this approach is not entirely assumption-free, it does not rely on the observational priors such as scaling relations, gas fractions, etc. that the previous studies require.

To date, few observational studies have placed direct constraints on the baryonic impact on the matter PS with WL measurements. HD15 applied their 15-parameter parameteric baryonic feedback models to the Canada France Hawaii Telescope Lensing Survey (CFHTLenS) cosmic shear data.
They found that the DMO model (zero neutrino mass and no baryonic feedback) is rejected at the $>2~\sigma$ level from their $p$-value test.
\cite{2017MNRAS.471.4412K} applied the HD15 model to the Kilo-Degree Survey (KiDS) 450 sq. degree data and 
and measured only the upper limit.
Most cosmology studies with KiDS \citep{10.1093/mnras/stw2805,2017MNRAS.471.1259J,2018MNRAS.474.4894J,2018MNRAS.476.4662V,2020A&A...633A..69H} used the M15 model to marginalize over the baryonic feedback effect, unable to  confine the feedback parameter posterior within their prior intervals.
In the cosmic shear studies with Hyper Suprime-Cam (HSC)  \citep{2019PASJ...71...43H,10.1093/pasj/psz138}, the results are consistent with the DMO case, presumably because of the conservative cuts in the cosmic shear measurements.

Based on the Deep Lens Survey (DLS) and the M15 model, \citet[][hereafter Y19]{2019ApJ...870..111Y} presented
the baryonic feedback measurement,
constraining both the lower and upper bounds of the 
feedback parameter.
Although the area is small ($\mytilde20$ sq. deg), the DLS depth is high, reaching down to $\mytilde 26.5^{th}$ in $B$, $V$, $R$, and $z$. This enables a competitive 
constraints on cosmological parameters (e.g., $S_8 = 0.810^{+0.039}_{-0.031}$)
compared to those of most Stage II and some early Stage III results. The DLS result is one of the few recent WL studies, whose measurements are highly consistent with the Planck value, $S_8 =0.832 \pm 0.013$ \citep{2018arXiv180706209P}.

However, it is difficult to interpret the Y19 result because taken at face value the measurement implies that the feedback strength should be much higher than the recipes in most state-of-the art hydrodynamical cosmological simulations.
As discussed in Y19, the insufficient degree of freedom in the M15 model may be one of the possible causes for this result. Also, we can consider possibilities that some other residual astrophysical systematics can masquerade as baryonic feedback. One such potentially relevant
astrophysical systematic error in Y19 is a non-linear
galaxy bias \citep{2020arXiv200407811A} because the measurement is obtained from the combination of galaxy clustering and galaxy-galaxy lensing under the assumption that the galaxy bias is linear at $l<2000$.

In this letter, we 
report measurement of baryonic feedback parameter from the KiDS-VIKING 450 sq. degree data (KV450). Specifically, we use the public data set used in \citet[][hereafter H20]{2020A&A...633A..69H}. 
Because the H20 study is based on cosmic shear, the nonlinear galaxy bias that may potentially have affected the Y19 result is not an issue in the current analysis. Also, as the KV450 WL pipeline is completely independent of the DLS one, consistent detection between the two different data sets serves as crucial consistency check.

\begin{figure*}[t]
\centering
    \includegraphics[ trim=0cm 0cm 0cm 1cm, width = 0.33\textwidth]{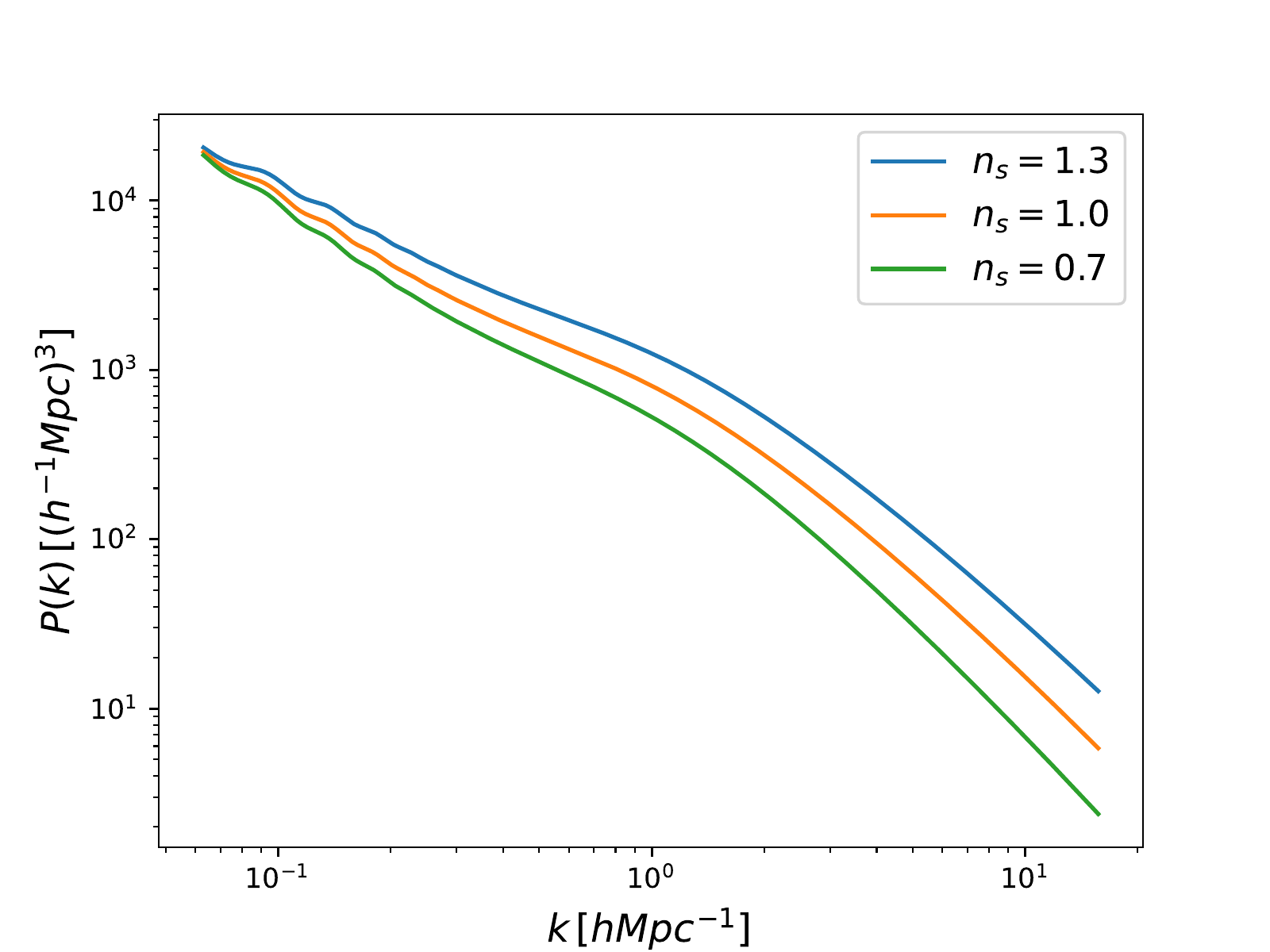}
    \includegraphics[trim=0cm 0cm 0cm 1cm, width = 0.33\textwidth]{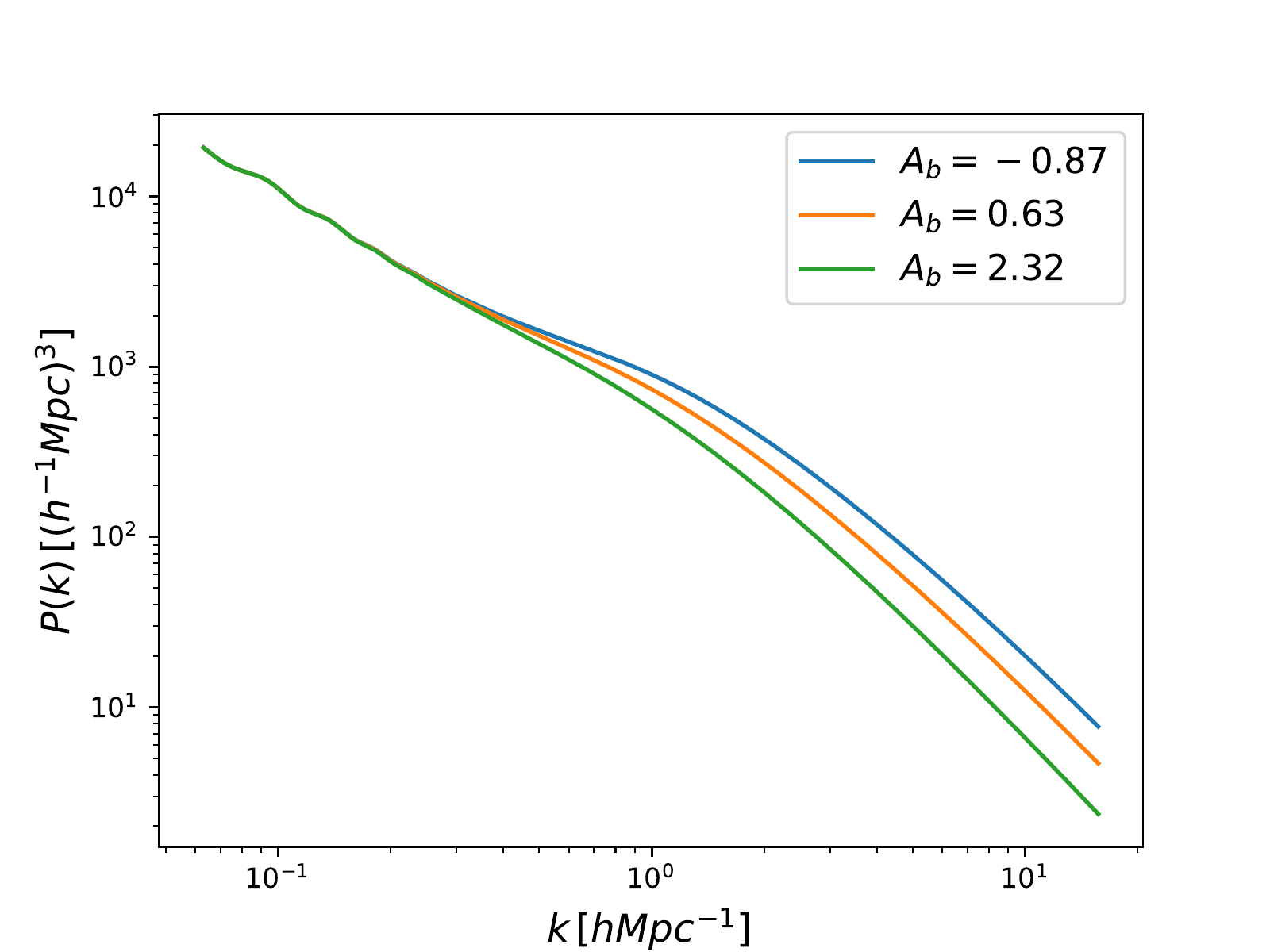}
    \includegraphics[trim=0cm 0cm -0.5cm 0.5cm, width = 0.26\textwidth, height=3.5cm]{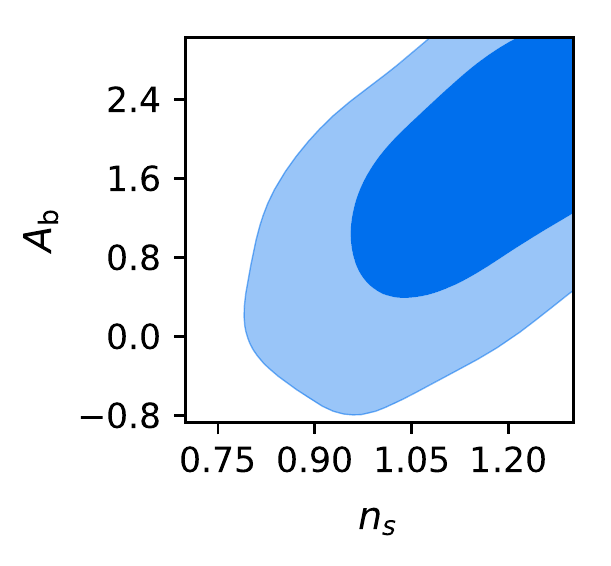}
\caption{Degeneracy between the spectral index $n_{\rm s}$ and baryonic feedback parameter $A_{\rm b}$.
The first and middle panels show the matter power spectra as a function of spectral index $n_{\rm s}$ and baryonic feedback parameter $A_{\rm b}$, respectively. The change in the matter PS at $ k \gtrsim 1~ \rm Mpc^{-1}$ due to the variation in $n_{\rm s} \in [0.7,1.3]$ is similar to the one due to the variation in $A_{\rm b} \in [-0.87, 2.32]$.
The third panel displays their posterior distributions.
The wide interval in $n_{\rm s}$ ($\in [0.7,1.3]$, encompassing the $\pm\mytilde75~\sigma$ range of the Planck result) substantially weakens the constraining power in $A_{\rm b}$. See \textsection~\ref{sec:prior_setting} for details.}

\label{fig:power_spectrum_ns_B}
\end{figure*}

\begin{figure}[t]
\centering
\includegraphics[width =0.48\textwidth]{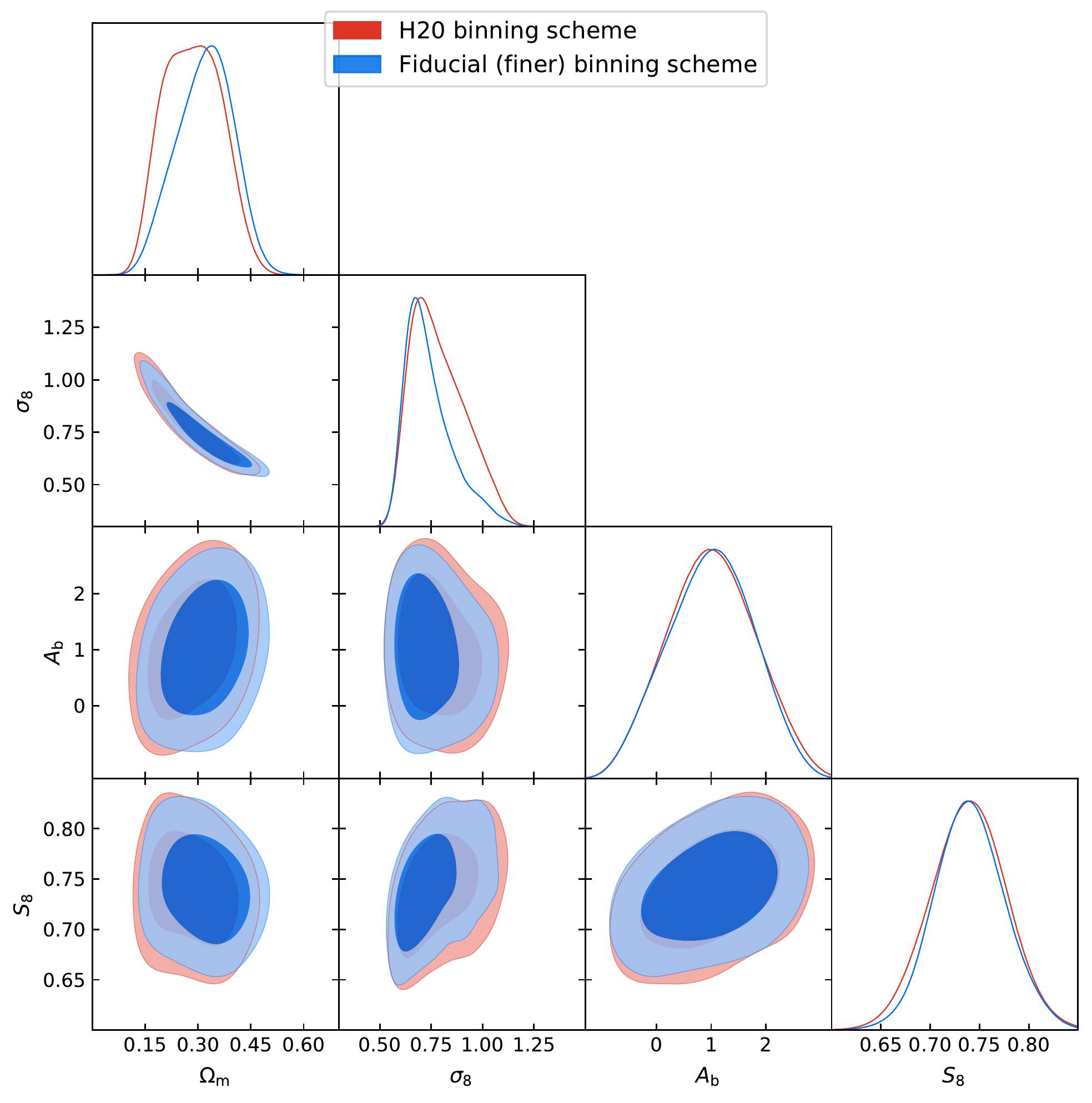}
\caption{Effect of angular binning scheme. The same priors were applied to both cases. We find that the constraining power of the H20 data improves for some cosmological parameters when we increase the number of angular bins from 7(6) (the original H20 binning scheme) bins to 10(8) bins for $\xi^+$($\xi^-$). The effect is most significant for the reduction of the degeneracy between $\Omega_m$ and $\sigma_8$; the uncertainty of $\sigma_8$ is reduced by $\mytilde20$\%.
The abruptly declining left tail of the $\Omega_m$ posterior for the H20 binning case is an artificial feature due to the employed kernel density estimator (smoothing) for plotting.
The uncertainty in $S_8$ decreases by $\mytilde7$\% while the impact on $A_{\rm b}$ is negligible. Readers are reminded that the result shown here with the H20 binning scheme (red) is our reproduction based the modified $n_s$ and $A_b$ priors, not identical to the one in the H20 paper.
} 
\label{fig:binning_effect}
\end{figure}

\section{Data}
\label{sec:data}
\defcitealias{2020A&A...633A..69H}{H20}
The details of  KV450 and its catalog used for cosmic shear analysis are described in \cite{2019A&A...632A..34W} and H20.
Below we only provide a brief description. The KV450 shape catalog was produced using {\tt lensfit} and the calibration methods described in \cite{2007MNRAS.382..315M,2013MNRAS.429.2858M,2019A&A...624A..92K,2017MNRAS.467.1627F}, while the photo-$z$ catalog was based on
{\tt BPZ} \citep{2000ApJ...536..571B} trained with the following spectroscopic samples: zCOSMOS \citep{2009ApJS..184..218L}, DEEP2 \citep{2013ApJS..208....5N}, VVDS \citep{2013A&A...559A..14L}, GAMA-G15Deep \citep{2018MNRAS.479.3746K}, and CDFS \citep{2013A&A...559A..14L,2008A&A...478...83V, 2016heas.confE..26V,  2013MNRAS.428.1281J}.

We use the same five tomographic binning schemes: $z_{\rm b} \in [0.1,0.3],\, [0.3,0.5],\, [0.5,0.7], \,[0.7,0.9], \mbox{and}\, [0.9,1.2]$ as in H20.
After confirming that our analysis pipeline reproduces the identical posteriors of H20 with the original data vectors and covariances, we choose to use finer angular binning for our subsequent analysis because we find that this increase in the number of angular bins improves the cosmological constraints for some parameters.
In H20, $\xi^+$($\xi^-$) were measured using 7(6) bins for the scale range [0.50, 72] ([4.2, 300]) arcmin uniformly divided in logarithmic scale.
We choose 10(8) bins for $\xi^+$($\xi^-$) over the similar angular range [0.50, 103]([4.2, 300]). Figure~\ref{fig:binning_effect} illustrates that the current binning scheme significantly reduces the parameter degeneracy between $\Omega_m$ and $\sigma_8$. For instance, while the full posterior of $\Omega_m$ in H20 is not contained within the prior interval, Figure~\ref{fig:binning_effect} shows that 
the peak, lower, and upper bounds are well-determined with our finer binning scheme. Also, 
the uncertainty of $\sigma_8$ is reduced by $\mytilde20$\%.

Note that we do not extend the lower angular limit of H20 to increase sensitivity to baryonic feedback because doing so will make our result also susceptible to other systematics such as nonlinear intrinsic alignment \citep{2020arXiv200302700F} and also because the impact of baryonic feedback at the smallest scale of H20 (0.5 and 5 arcmin in $\xi^+$ and $\xi^-$, respectively) is already significant (up to \mytilde20\% in the matter PS with respect to the DMO PS).

For our new binning scheme, shear-shear correlations were re-measured
using {\tt treecorr} (Jarvis et al. 2004). The corresponding covariance matrix was also re-calculated analytically with the same recipe as used/updated in Hildebrandt et al. (2016)/H20.

\section{Analysis}
\label{sec:analysis}
\subsection{Cosmology and Baryonic Feedback Models}

We compute the linear matter
PS with {\tt camb}\footnote{\url{http://camb.info}} \citep{Lewis:1999bs, Howlett:2012mh}. 
To account for the nonlinear evolution and baryonic feedback effects, we use the halo-model based code {\tt HMcode}\footnote{\url{https://github.com/alexander-mead/hmcode}} (M15). 
The ``halo model" approach is a significant improvement over the ``halofit" model \citep{2003MNRAS.341.1311S}, which requires many ($\mytilde38$) fitting parameters. This approach with more physically-motivated seven\footnote{The total number of free parameters is 14.} halo parameters can cover a wider range of cosmologies including different levels of baryonic feedback.
Also, this parameterization allows us to interpret the PS variation across different cosmological/feedback simulations in the astrophysical context.

The cosmology-dependent halo model parameters were determined 
using the power spectra derived from the {\small COSMIC EMU} \citep{2010ApJ...715..104H, Heitmann_2013}.
Using different baryonic feedback settings of OWLS,  
M15 find that among these seven halo parameters, the two parameters, namely the minimum halo concentration $A$
and the halo bloating factor $\eta$, need to be adjusted to accommodate the resulting change in the PS shape. In addition, although the number of the simulation sets is limited, they suggest that the two parameters are related as follows \citep{2018MNRAS.474.4894J}:
\begin{equation}
\eta_0 = 0.98 - 0.12 A, \label{eqn:Ab_eta}
\end{equation}
\noindent
where $\eta_0= \eta + 0.3 \sigma_8(z)$. 
The $A$ values of 2.32 and 3.13 correspond to
the AGN and DMO cases, respectively. This linear relation has been used/tested in many cosmological studies with different $A$ ranges  (e.g. [2.32, 3.13] in \cite{2019PASJ...71...43H}, [2, 3.13] in H20, [2, 4] in \cite{2018MNRAS.476.4662V} and \cite{10.1093/mnras/stw2805}, [2, 4]/[1, 10] in \cite{2017MNRAS.471.1259J}, and [1, 4]/ [1, 10] in \cite{2018MNRAS.474.4894J}).

In this study, we redefine the baryonic feedback parameter as follows:
\begin{equation}
A_{\rm b} \equiv 3.13 - A.
\end{equation}
\noindent
This definition makes a positive departure of $A_{\rm b}$ from zero (i.e., $A = 3.13$, DM-only case) mean positive feedback with a larger value corresponding to stronger feedback.

\subsection{Prior Settings}
\label{sec:prior_setting}

We use the same settings as in H20 in order to avoid potential confusion in interpretation and also enable a fair comparison of the resulting cosmological parameters.
Exceptions are made for the matter PS spectral index $n_{\rm s}$ and the baryonic feedback $A_{\rm b}$ parameters because of the reasons explained below.

As illustrated in Figure~\ref{fig:power_spectrum_ns_B}, the change in the matter PS at $ k \gtrsim 1~ \rm Mpc^{-1}$ due to the variation in $n_{\rm s} \in [0.7,1.3]$ is similar to the one due to the variation in $A_{\rm b} \in [-0.87,2.13]$.
This causes
a degeneracy between the two parameters
in their posterior distributions as shown in Figure~\ref{fig:power_spectrum_ns_B}.
Consequently, within the statistical noise level of KV450, it is difficult to distinguish their impacts. Since H20 used $n_{\rm s} \in [0.7,1.3]$, their $A_{\rm b}$ posterior is not bounded. We argue that 
the H20 $n_{\rm s}$ prior interval is too wide because it
corresponds to the $\pm \mytilde 75~\sigma$ range of the Planck constraint $n_\mathrm{s} = 0.965\pm 0.004$ \citep{2018arXiv180706209P} or $\pm \mytilde 21~\sigma$ range of the Nine-Year  Wilkinson  Microwave  Anisotropy (WMAP9) result $n_\mathrm{s} = 0.972\pm 0.013$ \citep{2013ApJS..208...19H}. 
Therefore, we choose to use 
the interval $n_{\rm s} \in [0.87,1.07]$ as our fiducial prior, which is still very conservative, corresponding to $\pm\mytilde25~(8)~\sigma$ range of the Planck (WMAP9) measurement. Similarly narrower $n_{\rm s}$ intervals are used by the Hyper-Suprime Cam surveys \citep{2019PASJ...71...43H}, Dark Energy Surveys Year 1 study \citep{2018PhRvD..98d3528T}, and DLS (Y19). Readers are reminded that the lower (upper) limit of the $h$ prior interval $[0.64,0.82]$ used in H20 and the current study corresponds to 6.8 (5.6) $\sigma$ of the direct (Planck) measurement \citep{2019ApJ...876...85R,2018arXiv180706209P}.

We extend the prior interval of $A_{\rm b}$ in the H20 setting because our initial experiment with the $A_{\rm b} \in [0.0,1.13]$ setting (H20) indicates that both tails of the posterior should exist outside this range. 
Thus, we use $A_{\rm b} \in [-0.87,3.03]$, which is also employed by Y19. 

\begin{figure}[t]
\centering
\includegraphics[ width = 0.43\textwidth]{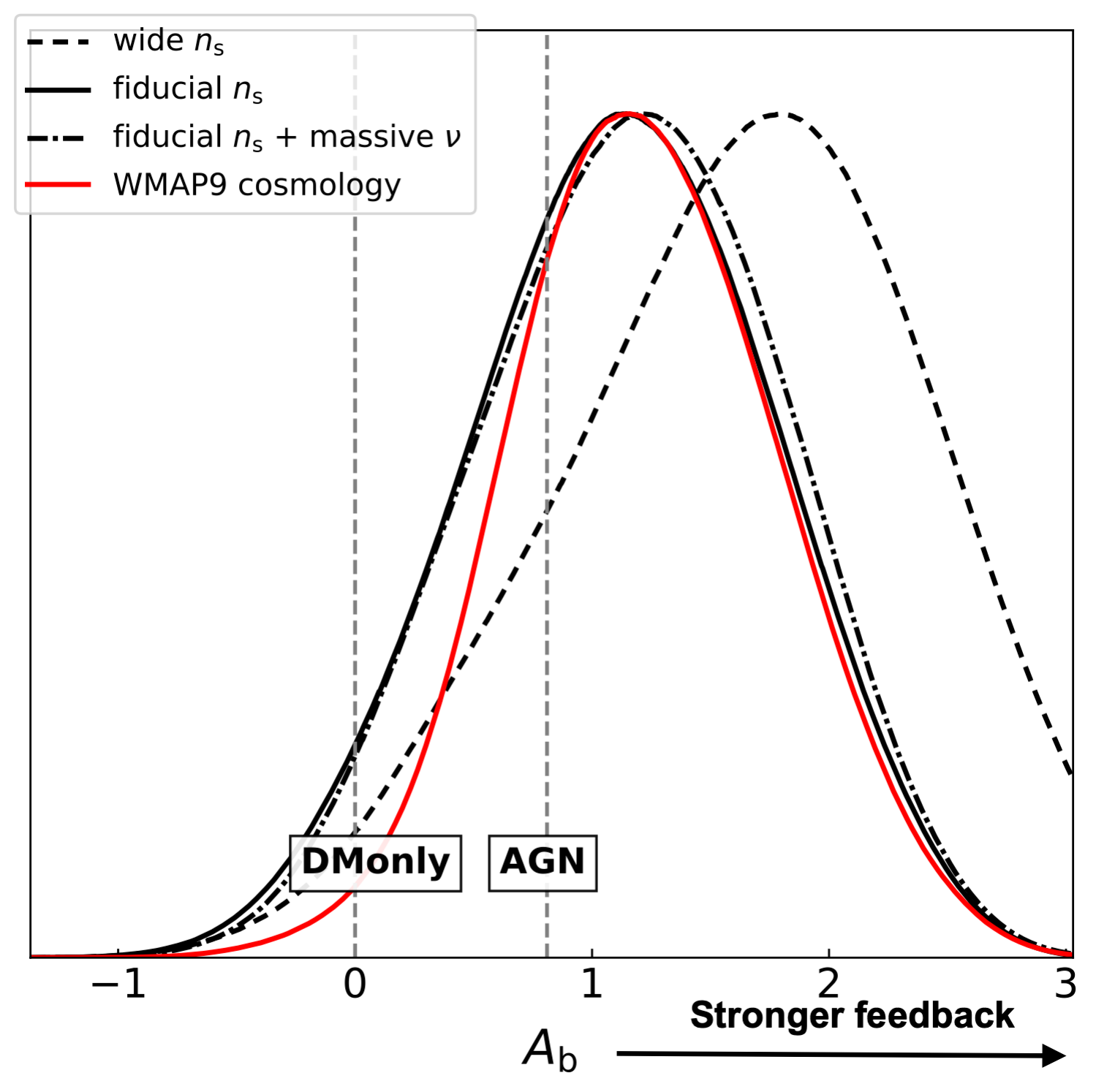}
\caption{Baryonic feedback constraints for different settings.
We determine both the lower and upper limits of the $A_{\rm b}$ posterior with the fiducial $n_{\rm s}$ prior whereas the use of the wide $n_{\rm s}$ interval does not constrain the upper bound.
Our fiducial measurement is consistent with the DM-only case at the 1.2~$\sigma$ level. Adding massive neutrinos virtually does not change the result.
When the WMAP9 cosmology is assumed, we exclude the DM-only scenario at the 2.2$\sigma$ level ($\mytilde98.5$\%, one-sided).
These measurements are summarized in Table~\ref{tab:B_constraint}.}  \label{fig:B_constraints}    
\end{figure}

\section{Result}
\label{sec:results}

\begin{table}\centering
\caption{Baryonic feedback and cosmological constraints \\ w/ different prior settings and external data combination.}
\scriptsize
\begin{tabular}{cccc}
\hline
\hline
Setting &  $A_{\rm b}$ & $S_8$ & $\chi^2/\mbox{d.o.f}^2$\\
\hline
wide $n_\mathrm{s}$ &$1.52_{-1.01}^{+0.78}$ &$0.745^{+0.041}_{-0.039}$&1.00\\
fiducial $n_\mathrm{s}$  & $1.01^{+0.80}_{-0.85}$
& $0.739^{+0.036}_{- 0.035}$&1.01
\\
\makecell{fiducial $n_\mathrm{s}$  + massive neutrino} & $1.07_{-0.86}^{+0.78}$& $0.737^{+0.035}_{-0.036}$ &1.02\\
\hline

WMAP9 cosmology & $1.21^{+0.61}_{-0.54}$ & Fixed & 0.99\\
Planck cosmology & $1.60^{+0.53}_{-0.52} $ & Fixed  & 1.01\\
\hline
 w/ SDSS DR12$^{1}$ & $1.29_{-0.82}^{+0.70} $ & $0.774 ^{+0.028}_{-0.032}$ & 1.01 \\
 w/ Planck TT likelihood & $1.32_{-0.66}^{+0.58}$ & $0.805^{+0.019}_{-0.019}$ &1.05\\
 w/ Planck TE likelihood & $0.74_{-0.72}^{+0.74} $ & $0.756^{+0.021}_{-0.018}$  & 1.02\\
 w/ Planck EE likelihood & $ 0.65_{-0.74}^{+0.76}$ & $0.746^{+0.027}_{-0.033} $  &1.02\\

\hline
\hline
\end{tabular}
\label{tab:B_constraint}
\tablecomments{The error bars denote the $\mytilde68.3$\% confidence intervals. 1. We used the baryonic acoustic oscillation and full shape measurements of redshift space distortion from the SDSS DR12 data \citep{2017MNRAS.470.2617A}. 2. The reduced $\chi^2$ values were evaluated at the best-fit locations for the cosmic shear data vector (i.e., we do not include the $\chi^2$ value for the external data).}
\end{table}

Our baryonic feedback measurements are summarized in Table~\ref{tab:B_constraint} for various settings. Note that the reduced $\chi^2$ values were evaluated at the best-fit locations in the parameter space only for the cosmic shear data vector. This investigation is to examine whether or not our constraints on $A_b$ are primarily due to tensions between the KV450 data and assumed (or preferred by external data) cosmology.
In all cases, the reduced $\chi^2$ values are turned out to be close to unity with small variations.

Figure~\ref{fig:B_constraints} displays the marginalized posterior of $A_b$ for four selected cases.
Our fiducial case ($0.87 \leq n_{\rm s} \leq 1.07$) with the KV450 data alone constrains both the lower and upper limits $1.01_{-0.85}^{+0.80}$. Since the ``wide" $n_{\rm s}$ ($0.7 \leq n_{\rm s} \leq 1.3$) setting cannot determine the upper limit within $A_{\rm b} \in [-0.87,3.03]$, the ambiguity in normalization prevents us from quantitatively interpreting the posterior.
The result from this ``wide" $n_{\rm s}$ case is consistent with the previous KiDS-450 3$\times$2pt result \citep{2018MNRAS.474.4894J}, which found a peak at $A_{\rm b} = 1.53$ with the 95\% lower bound at -0.17 (the upper bound is not constrained). 
Figure~\ref{fig:B_constraints} also shows that the result virtually remains the same when we include massive neutrinos ($0.06\leq \Sigma m_{\nu} \leq 0.9~\rm eV$). The fiducial KV450-only result is consistent at the $\mytilde1.2~\sigma$ level with the DMO case ($A_{\rm b}=0$). We confirmed that reducing the $n_{\rm s}$ prior interval further than this fiducial value does not significantly affect our $A_{\rm b}$ measurement. For instance, the choice of 
the Gaussian prior on $n_{\rm s}=0.965\pm0.004$ yields $A_b=0.97^{+0.76}_{-0.82}$.
The changes in $S_8\equiv \sigma_8 \sqrt{\Omega_m/0.3}$ and its uncertainty due to the $A_{\rm b}$ marginalization
are negligible in the KV450-only case (Table~1). 
H20 reports $S_8=0.737^{+0.040}_{-0.036}$ from the same data.
In our fiducial case, the shift in $S_8$ is only $\mytilde6$\% of the statistical uncertainty with little change in the measurement uncertainty.

The above measurement demonstrates that KV450 alone can put meaningful constraints on both cosmology and astrophysics simultaneously. However, a more common practice in astrophysics is to probe astrophysical properties with an assumption of a certain (fixed) cosmology. Here, we carry out such an experiment with the KV450 data. When we assume the cosmological parameters favored by the WMAP9 \citep{2013ApJS..208...19H} observation,
we obtain $A_{\rm b} = 1.21^{+0.61}_{-0.54}$. The shift in the central value is small with respect to the KV450-only case (see Figure~3) while the parameter uncertainty is reduced by $\mytilde37$\%. 
Therefore, this measurement excludes the DMO case (and thus detects the baryonic feedback) at the $\mytilde2.2\sigma$ level ($\mytilde98.5$\%, one-sided).
For the Planck  cosmology \citep{2018arXiv180706209P}, the central value of $A_{\rm b}$ increases by $\mytilde0.39$ compared to the WMAP9 case, favoring stronger feedback (see Table 1). This increase in $A_{\rm b}$ is by and large attributed to a significantly higher $S_8$ value favored by the Planck cosmology; stronger suppression is needed to reconcile the difference in $S_8$.
Interestingly, the reduced $\chi^2$ value (1.01) when the cosmology is fixed to the Planck result is in excellent agreement with the one in the fiducial case (Table 1). This potentially provides a chance to shed light on the origin of the much-discussed $S_8$ tension between WL and Planck results. When we examined the rest of the free parameters such as shear calibration, intrinsic alignment, photo-$z$ calibration, etc., we find that the changes are negligible. Thus, one naive interpretation is that the actual strength of the baryonic feedback might be stronger than the current predictions/measurements in other studies, as discussed in Yoon et al. (2019). However, we defer the issue to our future studies when more complete baryonic feedback model becomes available.

In addition to the above experiment with a fixed cosmology, one can also utilize external data to
constrain both cosmology and baryonic feedback.
We summarize these results in Table~1.
Combining KV450 with the redshift space distortion and baryonic acoustic oscillation measurement from the Sloan Digital Sky Survey Baryon Oscillation Spectroscopic Survey Data Release 12 \citep{2017MNRAS.470.2617A} yields $A_{\rm b} = 1.29^{+0.70}_{-0.82}$, which is in good agreement with the result obtained when we fix our cosmology to the WMAP9 result.
A similar result is obtained when the Planck TT data \citep{2019arXiv190712875P} are added. However, because this external data possess $\mytilde2~\sigma$ tension in $S_8$ with KV450, the interpretation should use caution. The combinations with the Planck TE and EE data, which do not present significant tensions with KV450, give consistent, but somewhat lower values.

\section{Discussion \& Summary}
\label{sec:discussion}

\begin{figure}[t] 
\centering
    \includegraphics[trim = 1.cm 0.6cm 1.4cm 1.5cm, width = 0.495\textwidth]{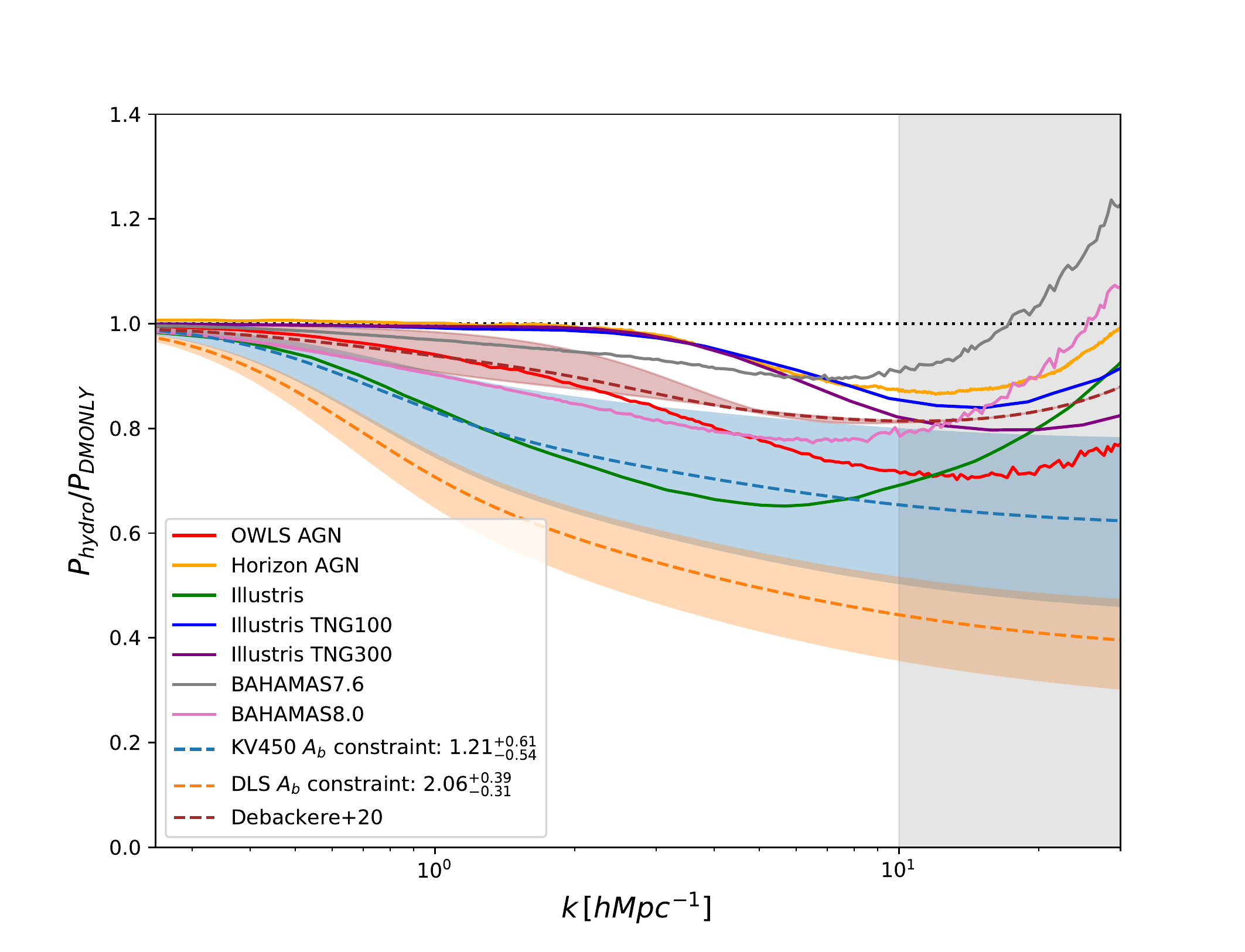}
\caption{PS ratio ($P_{\rm hydro}/P_{\rm DMonly}$) comparison at $z =0$.
We compare the M15 models constrained by the KV450 data (w/ WMAP9 cosmology) and DLS (w/ Planck combined) with the current state-of-the-art simulations: OWLS AGN, Horizon AGN \citep{2016MNRAS.463.3948D}, Illustris \citep{2015A&C....13...12N}, Illustris TNG100/300 \citep{2019ComAC...6....2N}, and BAHAMAS \citep{2017MNRAS.465.2936M}. 
The blue and orange shades represent the $1~\sigma$ uncertainties from KV450 and DLS, respectively, while the pink shade shows the hot gas model variation in Debackere et al. (2020).
With gray shade, we denote the power at $k \gtrsim 10 h \rm Mpc^{-1}$, whose impact on our WL data are negligible.}  
\label{fig:PS_simulation}    
\end{figure}

We have presented our baryonic feedback measurement from the KV450 data, constraining both the lower and upper limits of the feedback parameter.
H20 could not fully constrain this feedback parameter because the prior
interval in $A_{\rm b}$ is too narrow and because the prior interval in $n_{\rm s}$ is too wide. Readers are reminded that despite the changes of the prior intervals in these two parameters, the peak location of the $S_8$ posterior virtually remains unaffected, although its uncertainty reduces by $\mytilde8\%$.
Our best-fit value of $A_{\rm b}$ lies within the H20 prior range ($\in [0,1.13]$).

Our analysis with the KV450 data alone leads to the  feedback parameter measurement $A_{\rm b}=1.01^{+0.80}_{-0.85}$, which presents a consistency with the DMO case ($A_{\rm b}=0$) at the $\mytilde1.2~\sigma$ level. 
Under the assumption of the WMAP9 cosmology, we obtain $A_{\rm b}=1.21^{+0.61}_{-0.54}$. This result excludes the DMO case or provide evidence for baryonic feedback at the 2.2~$\sigma$ level ($\mytilde98.5$\%, one-sided).

Figure~\ref{fig:PS_simulation} illustrates
the level of the PS suppression at $z=0$ due to the baryonic feedback constrained from the current study. As mentioned above, our KV450 with the WMAP9 cosmology result is 
$\mytilde2.2\sigma$ away
from the DMO case $P_{hydro}/P_{DM }=1$.
At $k=10~h~\rm Mpc^{-1}$, the amount of the PS suppression is $\mytilde25$\% and the uncertainties encompass the OWLS-AGN, Illustris, and BAHAMAS 8.0\footnote{\url{http://powerlib.strw.leidenuniv.nl}} PS while
the DLS result (Y19) show some tensions with these predictions. Nevertheless, the current KV450 results are statistically consistent with the DLS ones.

A recent empirical study based on X-ray observations and halo occupation distribution modeling \citep{2020MNRAS.492.2285D} claims that 
a suppression at the $\mytilde15$\% level is expected at $k=5-10~h~\mbox{Mpc}^{-1}$.
This suppression is similar to the level predicted by the BAHAMAS 7.8 result and consistent with the current measurement at the $\mytilde1~\sigma$ level. Note that the prediction of \cite{2020MNRAS.492.2285D} is based on the WMAP9 cosmology.

\cite{2020arXiv2007.15633} presented a cosmic shear analysis with the KiDS 1000 sq. degree data. Marginalizing over the interval $A_{\rm b} \in [0.0,1.13]$ (i.e., $A_{\rm bary} \in [2,3.13]$ according to the notation in the paper), they obtained $A_{\rm b}=0.55^{-0.28}_{+0.39}$. As the authors noted, this is an artificial constraint because
the full posterior shape is not contained within this narrow prior range. Nevertheless, its peak location is fully consistent with the current measurement.

Just one day prior to the submission of the current paper, \cite{2020arXiv200715026H} uploaded their baryonic feedback measurement based on the Dark Energy Survey Year 1 (DES-Y1) data to the archive\footnote{http://https://arxiv.org/pdf/2007.15026.pdf}. Using the PCA approach and DES-Y1 data alone, they reported $Q_1 = 1.14^{+2.20}_{-2.80}$, where $Q_1$ is the first principal component amplitude. Although the left tail of the posterior is not well-determined with the DES-Y1 data alone, the result is consistent with the DMO case ($Q_1=0$). 
When combined with the Planck (EE+lowE) and BAO data, the constraint becomes
tighter $Q_1 = 1.42^{+1.63}_{-1.48}$, which is still consistent with the DMO case and excludes the most
extreme scenario in their comparison sample ($Q_1=5.84$) at the $\mytilde2$ sigma level.
Despite the difference in the characterization of the baryonic feedback strength, we find that the DES-Y1 measurement is fully consistent with ours, judging from their reference simulations compared with the posterior (Figure 15 of their paper).

Although we provide
the constraint on the baryonic feedback parameter from cosmic shear analysis alone, the proper interpretation should await improvements of our understanding in several aspects. First, as shown in Figure~\ref{fig:PS_simulation}, the current M15 one-parameter model lacks flexibility to accommodate the variation across different feedback scenarios. In particular,
the power on very small scales ($k\gtrsim10~h~\rm Mpc^{-1}$) does not show the ``upturns" due to baryonic cooling contraction, which however are present in most numerical studies.
Second, our nonlinear intrinsic alignment model is incomplete. The current model is based on the linear formalism \citep{2001MNRAS.320L...7C,2004PhRvD..70f3526H} with the replacement of the linear PS with the nonlinear one, which lacks solid theoretical justification. Nevertheless, given the statistical uncertainty of the KV450 data, we believe that perhaps the nonlinear effect is subdominant.

Despite the above caveats, however, the current study illustrates tremendous future opportunities that will be enabled with Stage IV WL surveys. 
Thanks to the unprecedented statistical powers, these studies will lead to precision cosmology as well as to a testbed for models beyond the standard $\Lambda$CDM (e.g., modified gravity, time-dependent dark energy, etc.). 
This can be empowered by understanding the small-scale systematics including baryonic feedback, which is a prerequisite to utilize the signals over wide scale ranges. Finally, in the future when the concordance cosmology is no longer in question, high S/N measurements of the suppressed PS shape
through gravitational lensing will provide critical feedback to obscure, sub-grid physics in numerical studies.

We thank Nora Elisa Chisari, Henk Hoekstra, Tilman Tr\"oster, Shahab Joudaki, Angus H. Wright, Alexander Mead, Hendrik Hildebrandt, and Catherine Heymans for carefully reading the manuscript and providing useful comments.
M. Yoon acknowledges support from the Max Planck Society and the Alexander von Humboldt Foundation in the framework of the Max Planck-Humboldt Research Award endowed by the Federal Ministry of Education and Research. M. Yoon also acknowledges support from the National Research Foundation of Korea (NRF) grant funded by the Korea government (MSIT) under no.2019R1C1C1010942.
M. J. Jee acknowledges support for the current research from the National Research Foundation of Korea under the program nos. 2017R1A2B2004644 and 2017R1A4A1015178.
\bibliographystyle{apj.bst}
\bibliography{main} 
\end{document}